\documentclass[a4paper,11pt]{article}
\usepackage{pos}
\usepackage{xspace}

\newcommand\Sgroup[2]{\mathrm{#1}(#2)}
\newcommand\su[1]{\ensuremath{\Sgroup{SU}{#1}}\xspace}
\newcommand\Nf{\ensuremath{N_{\mathrm{f}}}\xspace}
\newcommand\wzero{\ensuremath{w_0}\xspace}
\newcommand\stten{\ensuremath{\sqrt{\sigma}}\xspace}
\newcommand\mpcac{\ensuremath{m_{\mathrm{PCAC}}}\xspace}

\title{New lattice results for SU(2) gauge theory with one adjoint Dirac flavor}
\author*[a]{Ed Bennett}
\author[b,c]{Andreas Athenodorou}
\author[d]{Georg Bergner}
\author[e]{Biagio Lucini}

\affiliation[a]{Swansea Academy of Advanced Computing, Swansea University,\\
  Bay Campus, Swansea, UK}

\affiliation[b]{Dipartimento di Fisica, Universit\`a di Pisa and INFN, Sezione di Pisa, \\
  Largo Pontecorvo 3, 56127 Pisa, Italy}

\affiliation[c]{Computation-based Science and Technology Research Center, The Cyprus Institute,\\
20 Kavafi Str., Nicosia 2121, Cyprus}

\affiliation[d]{University of Jena, Institute for Theoretical Physics,\\
  Max-Wien-Platz 1, Jena, Germany}

\affiliation[e]{Department of Mathematics, Swansea University,\\
 Bay Campus, Swansea, UK}

\emailAdd{e.j.bennett@swansea.ac.uk}
\emailAdd{a.athenodorou@cyi.ac.cy}
\emailAdd{georg.bergner@uni-jena.de}
\emailAdd{b.lucini@swansea.ac.uk}

\abstract{Motivated by recent scenarios of exotic infrared behaviour and by earlier lattice findings, we present results for the \su{2} gauge theory with one Dirac flavor in the adjoint representation. This provides a major update on our previous investigation of this theory, including data for four values of the gauge coupling $\beta$, and for smaller masses and larger volumes than previously considered. Results for the particle spectrum, topological observables, and the anomalous dimension from both hyperscaling and the Dirac mode number are presented. At the finest coupling, we observe a mass anomalous dimension of $\gamma_* \gtrsim 0.6$. Our findings are analysed in relation to possible infrared behaviours of the model. In particular, we show that our results are not compatible with a confining scenario in which chiral symmetry is broken.
}

\FullConference{%
 The 38th International Symposium on Lattice Field Theory, LATTICE2021
  26th-30th July, 2021
  Zoom/Gather@Massachusetts Institute of Technology
}

\begin{document}
\maketitle

\section{Introduction}

The \su{2} theory with a single Dirac flavor in the adjoint representation has seen recent interest from a number of perspectives. The observation that the \su{2} theory with $\Nf=2$ adjoint Dirac flavors is conformal~\cite{Bursa:2011ru}, yet below the value of \Nf at which the two-loop beta function indicates the onset of the conformal window should lie \cite{Dietrich:2006cm}, raises the question of where the lower end of the conformal window actually lies. While the perturbative argument suggests that there should be no Banks--Zaks fixed point at $\Nf=1$, this must be tested from a first-principles non-perturbative analysis.

The possibility of a dynamical origin of electroweak symmetry breaking, and a composite Higgs boson originating from these strong dynamics, is another motivator behind interest in this family of theories. While the chiral symmetry breaking pattern in the one-flavor theory does not give sufficiently many Goldstone bosons to break electroweak symmetry, related models (for example, with one adjoint and two fundamental flavors \cite{Bergner:2021xx}) can overcome this limitation, and studying the $\Nf=1$ case gives a better understanding to build the more complex theories on.

Additionally, more recently the model has seen interest in the context of condensed matter theory, where it has been proposed that it may be dual to the critical theory describing the evolution of a specific topological phase transition \cite{Bi:2019gle}.

Previous work by the authors \cite{Athenodorou:2014eua} studied the mass spectrum and anomalous dimension of the theory at a single value of the coupling, observing a flavor-single scalar as the lightest state in the spectrum, approximately flat spectra when scaled by the string tension, and a large anomalous dimension. This was subsequently \cite{Athenodorou:2015fda} supplemented by an additional lighter value of the fermion mass, and a second value of the coupling, where a subset of the spectrum was observed, and the flat spectral ratios were seen to continue. More recently, Bi, et al.\ \cite{Bi:2019gle} have studied the theory at the same two values of the coupling, and observed the same spectrum, but additionally observed at a composite fermion state, whose mass overlaps with that of the spin-$\frac{1}{2}$ state studied in \cite{Athenodorou:2014eua}.

The observations to date have been interpreted as hinting towards the theory being near-conformal, but not yet conclusive. The range of fermion masses considered has been limited to relatively large values, and so it has not yet been possible to extract to the chiral limit. The continuum limit has also not yet been calculated; in particular, the value of the anomalous dimension was only observed a single lattice spacing, and so it has not yet been confirmed whether the observed large value persists at finer lattice spacings.

\section{Lattice setup}

We use the Wilson gauge action, and the Wilson fermion action, as detailed in \cite{Athenodorou:2021wom}. Since \su{2} is pseudoreal, and the adjoint representation is real, the action is entirely real, and we do not anticipate a sign problem.

We study four values of the gauge coupling $\beta=2.05,2.1,2.15,2.2$, and for each we study a range of fermion masses. The range of masses of the $2^{+}$ scalar baryon (which is measured in the $\gamma_5$ channel) is in the range (0.42, 1.08).

\section{Results}

In the following subsections, we present results for the mass spectrum, including finite-size hyperscaling and chiral perturbation theory analysis as well as details on the ratio of the tensor and scalar glueball masses, and the topological susceptibility. The methodology for each analysis is described in more detail in Ref.~\cite{Athenodorou:2021wom}, and all data are available in Ref.~\cite{datapackage}.

\subsection{Mass spectrum}

\begin{figure}
  \center
  \includegraphics[width=\textwidth]{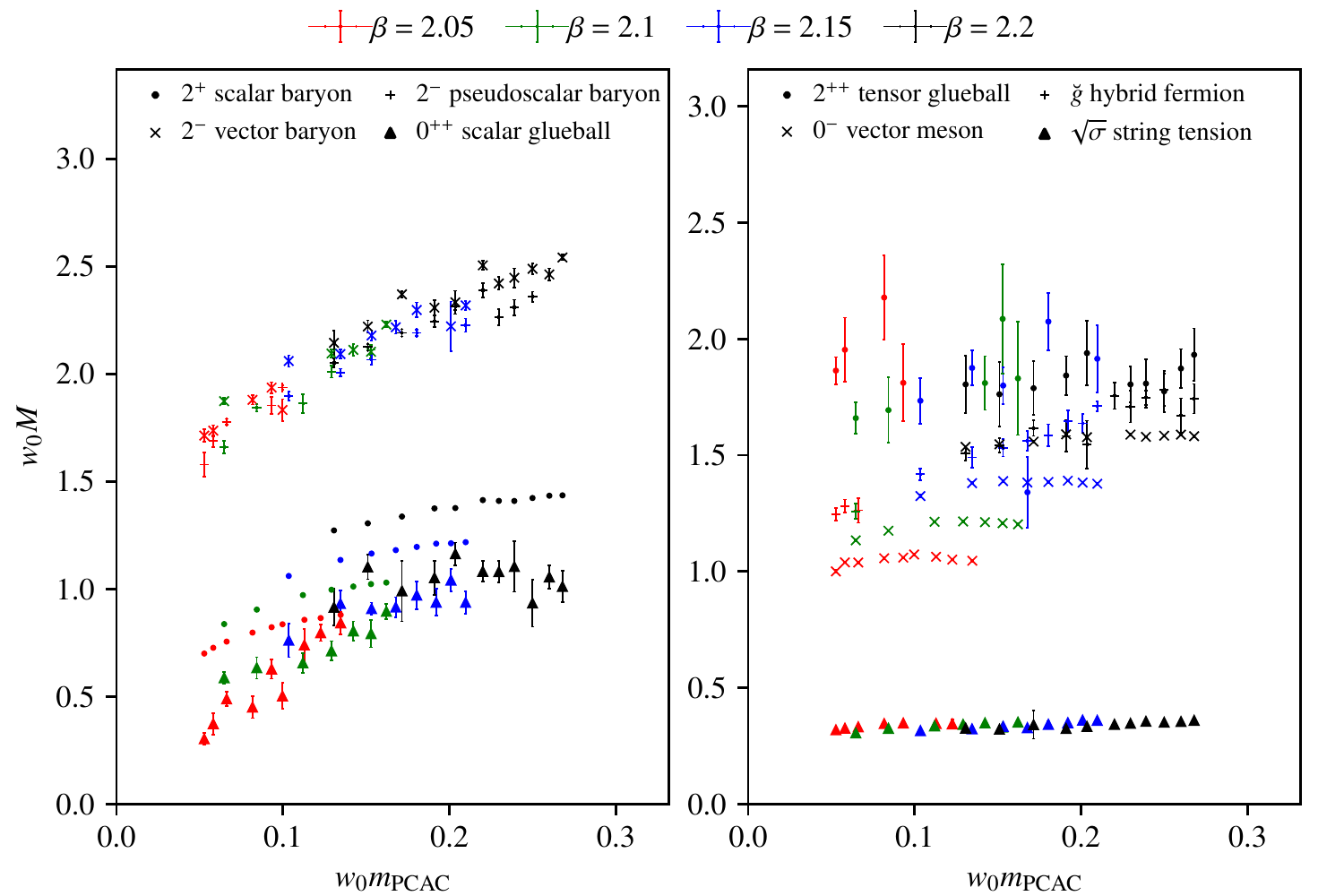}

  \caption{The mass spectrum of the theory as a function of the PCAC fermion mass, both scaled by the gradient flow scale \wzero.}\label{fig:spectrum}
\end{figure}

We measure the fermion mass via the partially conserved axial current (PCAC), and masses of mesons, baryons, glueballs, the hybrid spin-$\frac{1}{2}$ fermion state ($\breve{g}$), and the string tension \stten.

The observed mass spectrum of the theory is shown in Fig.~\ref{fig:spectrum}. In contrast to our previous work \cite{Athenodorou:2014eua,Athenodorou:2015fda}, where the spectrum was scaled by the string tension \stten, in this work we use the scale \wzero \cite{Borsanyi:2012zs} set by the Wilson flow \cite{Luscher:2010iy}. As \stten is observed to be constant as a function of \wzero for all $\beta$, the qualitative behavior of these ratios is expected to be the same. However, since \wzero is measurable more precisely than \stten, the uncertainties in the ratio are smaller than in our previous work.

This increased precision shows some deviation from the near-constant ratios observed in our previous work. In particular the $2^+$ scalar baryon increases as a function of the fermion mass, as does the $0^{++}$ scalar glueball in the $\beta=2.05$ case.

\subsection{Mass anomalous dimension}

\begin{figure}
  \center
  \includegraphics[width=\textwidth]{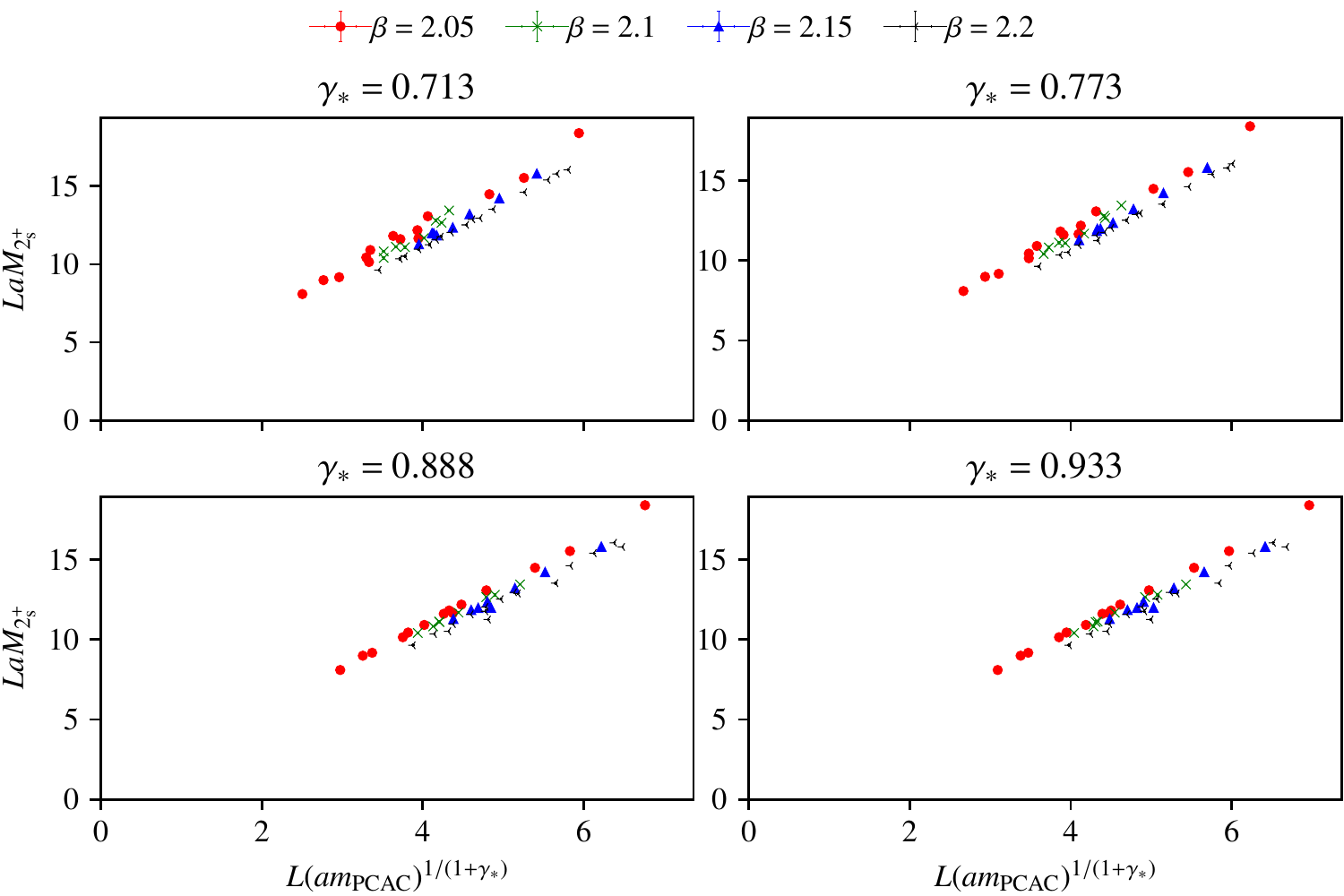}

  \caption{Results of the finite-size hyperscaling analysis. The four panels show the best fit values for the mass anomalous dimension at each value of $\beta$ studied.}\label{fig:fshs}
\end{figure}

The data for the mass of the $2^+$ scalar was analysed separately for each value of $\beta$, using a finite-size hyperscaling fit of the form
\begin{equation}
  L a M = f \left(L\left(a\mpcac\right)^{\frac{1}{1+\gamma_*}}\right)\;.
\end{equation}
As described in Ref.~\cite{Athenodorou:2021wom}, this was fitted using a global minimization of the sum of deviations from a set of local interpolating functions, which has the effect of pulling data from different values of the lattice extend $L$ together onto a common curve. Plots illustrating the quality of curve collapse for each $\beta$ are shown in Fig.~\ref{fig:fshs}.

The results indicate that as $\beta$ increases, the value of the mass anomalous dimension $\gamma_*$ decreases.

\subsection{Topological susceptibility}

\begin{figure}
  \center
  \includegraphics{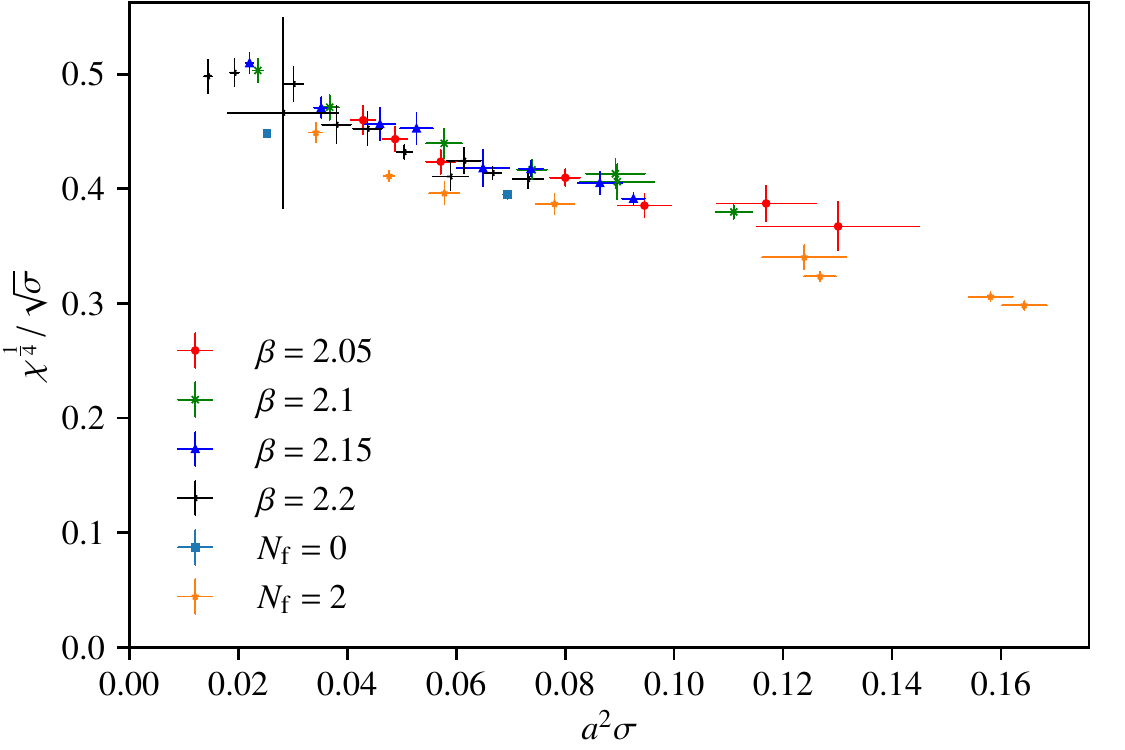}

  \caption{Results for the topological susceptibility normalised by the string tension, as a function of the string tension in lattice units. Results from Ref.~\cite{Bennett:2012ch} for the quenched and two-flavor theories are included for comparison.}\label{fig:susceptibility}
\end{figure}

The topological susceptibility is calculated from configurations smoothed using the Wilson flow. Following the argument presented in Ref.~\cite{Bennett:2012ch}, the topological susceptibility as a function of some scale for a mass-deformed conformal theory is expected to be the same as for the pure gauge theory. In this reference this has been confirmed for \su{2} with two adjoint Dirac flavors.

Results for the topological susceptibility of the $\Nf=1$ theory are presented in Fig.~\ref{fig:susceptibility}, where data from Ref.~\cite{Bennett:2012ch} are included for comparison. The data lie close to the pure gauge data, but are systematically slightly higher.

\subsection{Chiral perturbation theory}

\begin{figure}
  \center
  \includegraphics{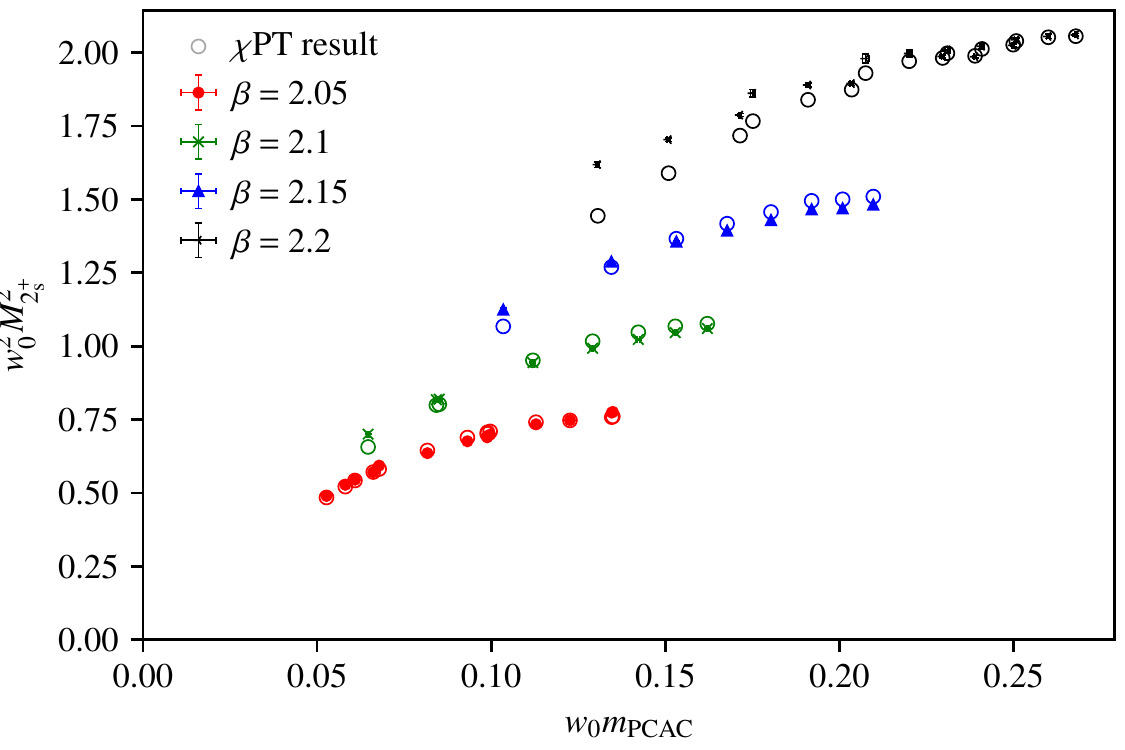}

  \caption{Results of a fit of the $2^+$ scalar baryon mass to the chiral perturbation theory ansatz in the text.}\label{fig:Xpt}
\end{figure}

To contrast the hyperscaling fit performed above, we also fit the data for the $2^+$ scalar baryon using a chiral perturbation theory ansatz of the form
\begin{equation}
  M^2 w_0^2 = 2B\hat{m}\left(1+L\hat{m}+D_1\hat{m}\log(D_2 \hat{m})\right)+W_1 a\mpcac + W_2 \left(\frac{w_0}{a}\right)^{-2}\;,
\end{equation}
where $M$ is the $2^+$ scalar baryon mass, and $\hat{m}=\mpcac w_0$. The results of this fit are shown in Fig.~\ref{fig:Xpt}. Tension is visible between the fit form and the data at all values of $\beta$, but this is most pronounced at higher values of $\beta$.

Since Fig.~\ref{fig:spectrum} shows qualititavely different behavior for $\beta=2.05$, we considered that data from this $\beta$ could be anchoring the fit, and thus causing the higher values of $\beta$ to be poorly described by the fit result. To exclude this scenario we repeated the fit, excluding the $\beta=2.05$ data; the results found were qualitatively very similar to the fit results shown.

\subsection{Tensor--scalar mass ratio}

\begin{figure}
  \center
  \includegraphics{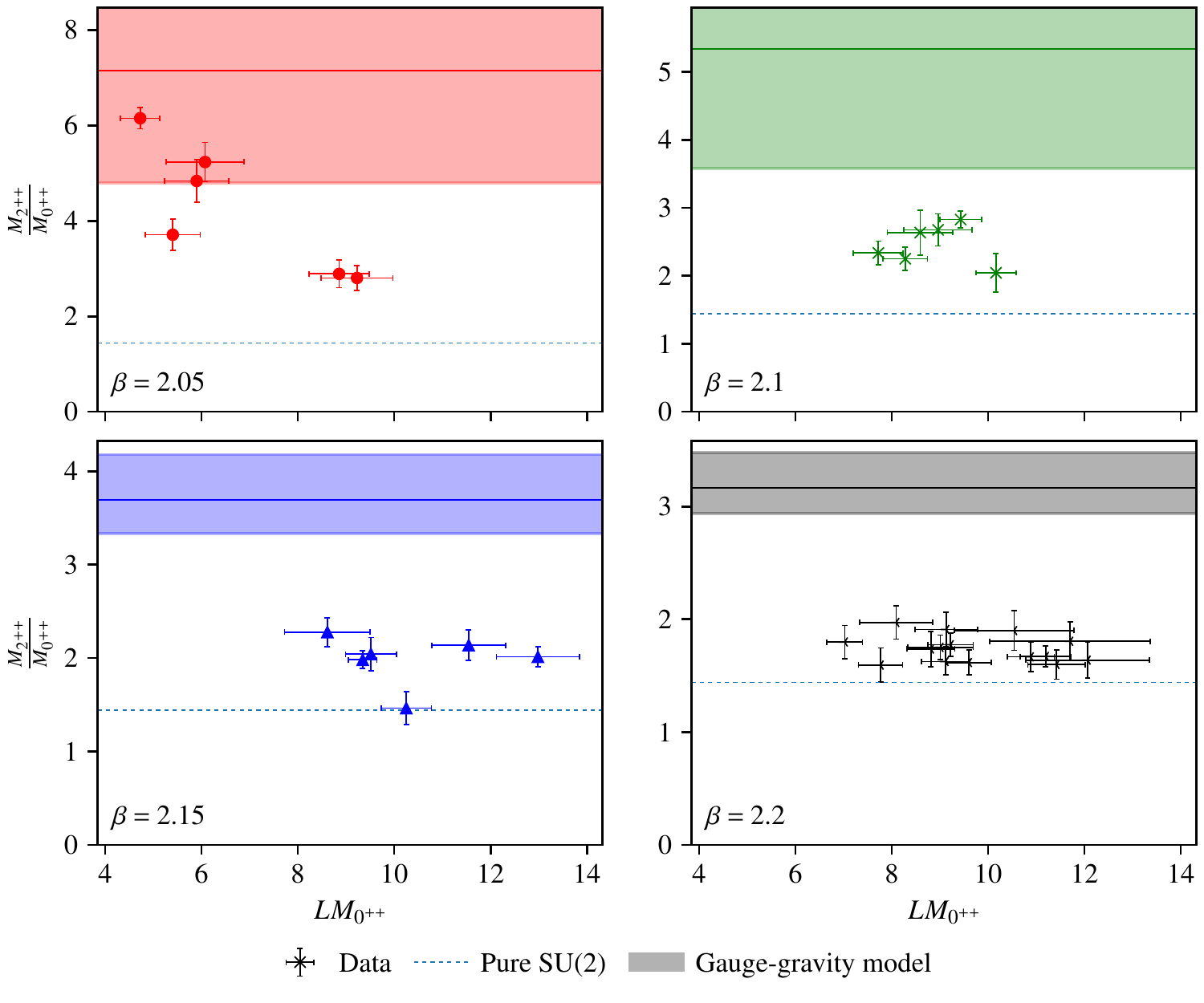}

  \caption{The ratio of the $2^{++}$ glueball mass to that of the $0^{++}$ as a function of the $0^{++}$ glueball mass scaled by the lattice extent $L$ for the four values of $\beta$ considered. The pure gauge value and the prediction from gauge--gravity duality are indicated for each case.}\label{fig:R_ratio}
\end{figure}

In Ref.~\cite{Athenodorou:2016ndx}, we proposed that the ratio of masses of the tensor and scalar glueballs may obey some universal behavior between a variety of disparate theories. The value of this ratio for the ensembles considered in this work are shown in Fig.~\ref{fig:R_ratio}. Also included in the plots are the values of $R$ in the pure gauge \su{2} case, and in the gauge--gravity model discussed in Ref.~\cite{Athenodorou:2016ndx} for the values of the anomalous dimension observed in the finite-size hyperscaling analysis for each choice of $\beta$.

The results in most cases lie between the two predictions, with higher values of $\beta$ lying closer to the pure gauge case, and only the $\beta=2.05$ case reaching the region predicted by the gauge--gravity model. This does not conclusively disprove the conjective in Ref.~\cite{Athenodorou:2016ndx}, as the values of $L$ and $am$ considered in this work were not specifically chosen to enable this analysis, and the behaviour as a function of $L$ and of \mpcac are difficult to disentangle. We anticipate that a dedicated study with specifically-chosen values of $L$ and $am$ would allow more concrete conclusions to be drawn.

\section{Conclusions}

We have studied the \su{2} gauge theory with one flavor of Dirac fermion in the adjoint representation at four values of the lattice spacing. In the spectrum, we observe a flavor-singlet scalar as the lightest state, and most spectral ratios are flat. The results for each $\beta$ are consistent with finite-size hyperscaling, but the mass anomalous dimension required for this consistency decreases as the value of $\beta$ increases. The data do not fit well to a chiral perturbation theory ansatz, in particular at larger values of $\beta$. The topological susceptibility shows similar behavior to that observed in the pure gauge and two-flavor case as a function of the string tension, and has the same gradient, but lies systematically slightly above the other theories.

We identify three possible interpretations of these results:

\begin{enumerate}
\item The theory is QCD-like in the infrared with broken chiral symmetry, and the large deviations in the chiral perturbation theory fits a result of lattice artefacts. This appears to be contradicted by the lightness of the scalar state; however, this could in principle be a result of being insufficiently close to the chiral limit.
\item The theory is conformal, but with scaling deviations causing the change in anomalous dimension. We are currently unable to reliably determine these violations, and it is not clear which value of $\beta$ is most reliable and closest to a conformal fixed point.
\item The theory is conformal, but with a stronger influence from the bulk phase than anticipated. Our earlier work \cite{Athenodorou:2014eua} took care to simulate above the onset of the bulk phase, however it is possible that some influence of this remains. In this case, the largest value of $\beta$ would be expected to be least affected by this effect, and so the smaller value of the mass anomalous dimension is more accurate.
\end{enumerate}

To confirm or rule out the first interpretation, it is necessary to simulate closer to the chiral limit. To probe the latter two interpretations, instead larger values of $\beta$ must be studied to enable a stable continuum limit extrapolation. Work on both directions is ongoing.

\begin{acknowledgments}
    We thank M.~Piai and D.~Elander for providing the data on the $R=M_{2^{++}}/M_{0^{++}}$ for the string inspired toy model. We also thank M.~\"Unsal for discussions.
   AA has been financially supported by the European Union's Horizon 2020 research and innovation programme ``Tips in SCQFT'' under the Marie Sk\l odowska-Curie grant agreement No.~791122. The work of EB has been funded by the Supercomputing Wales project, which is part-funded by the European Regional Development Fund (ERDF) via Welsh Government, by the Japan Society for the Promotion of Science under award PE13578, and by the UKRI Science and Technologies Facilities Council (STFC) Research Software Engineering Fellowship EP/V052489/1.
   The work of BL has been supported in part by the European Research Council (ERC) under the European Union’s Horizon 2020 research and innovation programme under grant agreement No 813942 and by the UKRI STFC Consolidated Grant ST/P00055X/1. The work of BL is further supported in part by the Royal Society Wolfson Research Merit Award WM170010 and by the Leverhulme Foundation Research Fellowship RF-2020-461\textbackslash \!9. GB acknowledges support from the Deutsche Forschungsgemeinschaft (DFG) Grant No.~BE 5942/3-1. Numerical simulations have been performed on the SUNBIRD system in Swansea, part of the Supercomputing Wales project, which is part-funded by the ERDF via Welsh Government.
\end{acknowledgments}

\bibliographystyle{apsrev4-2.bst}
\bibliography{references}

%apsrev4-2.bst 2019-01-14 (MD) hand-edited version of apsrev4-1.bst
%Control: key (0)
%Control: author (72) initials jnrlst
%Control: editor formatted (1) identically to author
%Control: production of article title (-1) disabled
%Control: page (0) single
%Control: year (1) truncated
%Control: production of eprint (0) enabled
\begin{thebibliography}{12}%
\makeatletter
\providecommand \@ifxundefined [1]{%
 \@ifx{#1\undefined}
}%
\providecommand \@ifnum [1]{%
 \ifnum #1\expandafter \@firstoftwo
 \else \expandafter \@secondoftwo
 \fi
}%
\providecommand \@ifx [1]{%
 \ifx #1\expandafter \@firstoftwo
 \else \expandafter \@secondoftwo
 \fi
}%
\providecommand \natexlab [1]{#1}%
\providecommand \enquote  [1]{``#1''}%
\providecommand \bibnamefont  [1]{#1}%
\providecommand \bibfnamefont [1]{#1}%
\providecommand \citenamefont [1]{#1}%
\providecommand \href@noop [0]{\@secondoftwo}%
\providecommand \href [0]{\begingroup \@sanitize@url \@href}%
\providecommand \@href[1]{\@@startlink{#1}\@@href}%
\providecommand \@@href[1]{\endgroup#1\@@endlink}%
\providecommand \@sanitize@url [0]{\catcode `\\12\catcode `\$12\catcode
  `\&12\catcode `\#12\catcode `\^12\catcode `\_12\catcode `\%12\relax}%
\providecommand \@@startlink[1]{}%
\providecommand \@@endlink[0]{}%
\providecommand \url  [0]{\begingroup\@sanitize@url \@url }%
\providecommand \@url [1]{\endgroup\@href {#1}{\urlprefix }}%
\providecommand \urlprefix  [0]{URL }%
\providecommand \Eprint [0]{\href }%
\providecommand \doibase [0]{https://doi.org/}%
\providecommand \selectlanguage [0]{\@gobble}%
\providecommand \bibinfo  [0]{\@secondoftwo}%
\providecommand \bibfield  [0]{\@secondoftwo}%
\providecommand \translation [1]{[#1]}%
\providecommand \BibitemOpen [0]{}%
\providecommand \bibitemStop [0]{}%
\providecommand \bibitemNoStop [0]{.\EOS\space}%
\providecommand \EOS [0]{\spacefactor3000\relax}%
\providecommand \BibitemShut  [1]{\csname bibitem#1\endcsname}%
\let\auto@bib@innerbib\@empty
%</preamble>
\bibitem [{\citenamefont {Bursa}\ \emph {et~al.}(2011)\citenamefont {Bursa},
  \citenamefont {Del~Debbio}, \citenamefont {Henty}, \citenamefont {Kerrane},
  \citenamefont {Lucini}, \citenamefont {Patella}, \citenamefont {Pica},
  \citenamefont {Pickup},\ and\ \citenamefont {Rago}}]{Bursa:2011ru}%
  \BibitemOpen
  \bibfield  {author} {\bibinfo {author} {\bibfnamefont {F.}~\bibnamefont
  {Bursa}}, \bibinfo {author} {\bibfnamefont {L.}~\bibnamefont {Del~Debbio}},
  \bibinfo {author} {\bibfnamefont {D.}~\bibnamefont {Henty}}, \bibinfo
  {author} {\bibfnamefont {E.}~\bibnamefont {Kerrane}}, \bibinfo {author}
  {\bibfnamefont {B.}~\bibnamefont {Lucini}}, \bibinfo {author} {\bibfnamefont
  {A.}~\bibnamefont {Patella}}, \bibinfo {author} {\bibfnamefont
  {C.}~\bibnamefont {Pica}}, \bibinfo {author} {\bibfnamefont {T.}~\bibnamefont
  {Pickup}},\ and\ \bibinfo {author} {\bibfnamefont {A.}~\bibnamefont {Rago}},\
  }\href {https://doi.org/10.1103/PhysRevD.84.034506} {\bibfield  {journal}
  {\bibinfo  {journal} {Phys. Rev. D}\ }\textbf {\bibinfo {volume} {84}},\
  \bibinfo {pages} {034506} (\bibinfo {year} {2011})},\ \Eprint
  {https://arxiv.org/abs/1104.4301} {arXiv:1104.4301 [hep-lat]} \BibitemShut
  {NoStop}%
\bibitem [{\citenamefont {Dietrich}\ and\ \citenamefont
  {Sannino}(2007)}]{Dietrich:2006cm}%
  \BibitemOpen
  \bibfield  {author} {\bibinfo {author} {\bibfnamefont {D.~D.}\ \bibnamefont
  {Dietrich}}\ and\ \bibinfo {author} {\bibfnamefont {F.}~\bibnamefont
  {Sannino}},\ }\href {https://doi.org/10.1103/PhysRevD.75.085018} {\bibfield
  {journal} {\bibinfo  {journal} {Phys. Rev. D}\ }\textbf {\bibinfo {volume}
  {75}},\ \bibinfo {pages} {085018} (\bibinfo {year} {2007})},\ \Eprint
  {https://arxiv.org/abs/hep-ph/0611341} {arXiv:hep-ph/0611341} \BibitemShut
  {NoStop}%
\bibitem [{\citenamefont {Bergner}\ and\ \citenamefont
  {Piemonte}(2021)}]{Bergner:2021xx}%
  \BibitemOpen
  \bibfield  {author} {\bibinfo {author} {\bibfnamefont {G.}~\bibnamefont
  {Bergner}}\ and\ \bibinfo {author} {\bibfnamefont {S.}~\bibnamefont
  {Piemonte}},\ }in\ \href@noop {} {\emph {\bibinfo {booktitle} {{38th
  International Symposium on Lattice Field Theory}}}}\ (\bibinfo {year}
  {2021})\BibitemShut {NoStop}%
\bibitem [{\citenamefont {Bi}\ \emph {et~al.}(2019)\citenamefont {Bi},
  \citenamefont {Grebe}, \citenamefont {Kanwar}, \citenamefont {Ledwith},
  \citenamefont {Murphy},\ and\ \citenamefont {Wagman}}]{Bi:2019gle}%
  \BibitemOpen
  \bibfield  {author} {\bibinfo {author} {\bibfnamefont {Z.}~\bibnamefont
  {Bi}}, \bibinfo {author} {\bibfnamefont {A.}~\bibnamefont {Grebe}}, \bibinfo
  {author} {\bibfnamefont {G.}~\bibnamefont {Kanwar}}, \bibinfo {author}
  {\bibfnamefont {P.}~\bibnamefont {Ledwith}}, \bibinfo {author} {\bibfnamefont
  {D.}~\bibnamefont {Murphy}},\ and\ \bibinfo {author} {\bibfnamefont {M.~L.}\
  \bibnamefont {Wagman}},\ }\href {https://doi.org/10.22323/1.363.0127}
  {\bibfield  {journal} {\bibinfo  {journal} {PoS}\ }\textbf {\bibinfo {volume}
  {LATTICE2019}},\ \bibinfo {pages} {127} (\bibinfo {year} {2019})},\ \Eprint
  {https://arxiv.org/abs/1912.11723} {arXiv:1912.11723 [hep-lat]} \BibitemShut
  {NoStop}%
\bibitem [{\citenamefont {Athenodorou}\ \emph {et~al.}(2015)\citenamefont
  {Athenodorou}, \citenamefont {Bennett}, \citenamefont {Bergner},\ and\
  \citenamefont {Lucini}}]{Athenodorou:2014eua}%
  \BibitemOpen
  \bibfield  {author} {\bibinfo {author} {\bibfnamefont {A.}~\bibnamefont
  {Athenodorou}}, \bibinfo {author} {\bibfnamefont {E.}~\bibnamefont
  {Bennett}}, \bibinfo {author} {\bibfnamefont {G.}~\bibnamefont {Bergner}},\
  and\ \bibinfo {author} {\bibfnamefont {B.}~\bibnamefont {Lucini}},\ }\href
  {https://doi.org/10.1103/PhysRevD.91.114508} {\bibfield  {journal} {\bibinfo
  {journal} {Phys. Rev. D}\ }\textbf {\bibinfo {volume} {91}},\ \bibinfo
  {pages} {114508} (\bibinfo {year} {2015})},\ \Eprint
  {https://arxiv.org/abs/1412.5994} {arXiv:1412.5994 [hep-lat]} \BibitemShut
  {NoStop}%
\bibitem [{\citenamefont {Athenodorou}\ \emph {et~al.}(2017)\citenamefont
  {Athenodorou}, \citenamefont {Bennett}, \citenamefont {Bergner},\ and\
  \citenamefont {Lucini}}]{Athenodorou:2015fda}%
  \BibitemOpen
  \bibfield  {author} {\bibinfo {author} {\bibfnamefont {A.}~\bibnamefont
  {Athenodorou}}, \bibinfo {author} {\bibfnamefont {E.}~\bibnamefont
  {Bennett}}, \bibinfo {author} {\bibfnamefont {G.}~\bibnamefont {Bergner}},\
  and\ \bibinfo {author} {\bibfnamefont {B.}~\bibnamefont {Lucini}},\ }\href
  {https://doi.org/10.1142/S0217751X17470066} {\bibfield  {journal} {\bibinfo
  {journal} {Int. J. Mod. Phys. A}\ }\textbf {\bibinfo {volume} {32}},\
  \bibinfo {pages} {1747006} (\bibinfo {year} {2017})},\ \Eprint
  {https://arxiv.org/abs/1507.08892} {arXiv:1507.08892 [hep-lat]} \BibitemShut
  {NoStop}%
\bibitem [{\citenamefont {Athenodorou}\ \emph
  {et~al.}(2021{\natexlab{a}})\citenamefont {Athenodorou}, \citenamefont
  {Bennett}, \citenamefont {Bergner},\ and\ \citenamefont
  {Lucini}}]{Athenodorou:2021wom}%
  \BibitemOpen
  \bibfield  {author} {\bibinfo {author} {\bibfnamefont {A.}~\bibnamefont
  {Athenodorou}}, \bibinfo {author} {\bibnamefont {Bennett}}, \bibinfo {author}
  {\bibfnamefont {G.}~\bibnamefont {Bergner}},\ and\ \bibinfo {author}
  {\bibfnamefont {B.}~\bibnamefont {Lucini}},\ }\Eprint
  {https://arxiv.org/abs/2103.10485} {arXiv:2103.10485 [hep-lat]}  (\bibinfo
  {year} {2021}{\natexlab{a}})\BibitemShut {NoStop}%
\bibitem [{\citenamefont {Athenodorou}\ \emph
  {et~al.}(2021{\natexlab{b}})\citenamefont {Athenodorou}, \citenamefont
  {Bennett}, \citenamefont {Bergner},\ and\ \citenamefont
  {Lucini}}]{datapackage}%
  \BibitemOpen
  \bibfield  {author} {\bibinfo {author} {\bibfnamefont {A.}~\bibnamefont
  {Athenodorou}}, \bibinfo {author} {\bibfnamefont {E.}~\bibnamefont
  {Bennett}}, \bibinfo {author} {\bibfnamefont {G.}~\bibnamefont {Bergner}},\
  and\ \bibinfo {author} {\bibfnamefont {B.}~\bibnamefont {Lucini}},\ }\href
  {https://doi.org/10.5281/zenodo.5139618} {\bibinfo {title} {Investigating the
  conformal behaviour of su(2) with one adjoint dirac flavor --- data release}}
  (\bibinfo {year} {2021}{\natexlab{b}})\BibitemShut {NoStop}%
\bibitem [{\citenamefont {Borsanyi}\ \emph {et~al.}(2012)\citenamefont
  {Borsanyi} \emph {et~al.}}]{Borsanyi:2012zs}%
  \BibitemOpen
  \bibfield  {author} {\bibinfo {author} {\bibfnamefont {S.}~\bibnamefont
  {Borsanyi}} \emph {et~al.},\ }\href {https://doi.org/10.1007/JHEP09(2012)010}
  {\bibfield  {journal} {\bibinfo  {journal} {JHEP}\ }\textbf {\bibinfo
  {volume} {09}},\ \bibinfo {pages} {010}},\ \Eprint
  {https://arxiv.org/abs/1203.4469} {arXiv:1203.4469 [hep-lat]} \BibitemShut
  {NoStop}%
\bibitem [{\citenamefont {L\"uscher}(2010)}]{Luscher:2010iy}%
  \BibitemOpen
  \bibfield  {author} {\bibinfo {author} {\bibfnamefont {M.}~\bibnamefont
  {L\"uscher}},\ }\href {https://doi.org/10.1007/JHEP08(2010)071} {\bibfield
  {journal} {\bibinfo  {journal} {JHEP}\ }\textbf {\bibinfo {volume} {08}},\
  \bibinfo {pages} {071}},\ \bibinfo {note} {[Erratum: JHEP 03, 092 (2014)]},\
  \Eprint {https://arxiv.org/abs/1006.4518} {arXiv:1006.4518 [hep-lat]}
  \BibitemShut {NoStop}%
\bibitem [{\citenamefont {Bennett}\ and\ \citenamefont
  {Lucini}(2013)}]{Bennett:2012ch}%
  \BibitemOpen
  \bibfield  {author} {\bibinfo {author} {\bibfnamefont {E.}~\bibnamefont
  {Bennett}}\ and\ \bibinfo {author} {\bibfnamefont {B.}~\bibnamefont
  {Lucini}},\ }\href {https://doi.org/10.1140/epjc/s10052-013-2426-6}
  {\bibfield  {journal} {\bibinfo  {journal} {Eur. Phys. J. C}\ }\textbf
  {\bibinfo {volume} {73}},\ \bibinfo {pages} {2426} (\bibinfo {year}
  {2013})},\ \Eprint {https://arxiv.org/abs/1209.5579} {arXiv:1209.5579
  [hep-lat]} \BibitemShut {NoStop}%
\bibitem [{\citenamefont {Athenodorou}\ \emph {et~al.}(2016)\citenamefont
  {Athenodorou}, \citenamefont {Bennett}, \citenamefont {Bergner},
  \citenamefont {Elander}, \citenamefont {Lin}, \citenamefont {Lucini},\ and\
  \citenamefont {Piai}}]{Athenodorou:2016ndx}%
  \BibitemOpen
  \bibfield  {author} {\bibinfo {author} {\bibfnamefont {A.}~\bibnamefont
  {Athenodorou}}, \bibinfo {author} {\bibfnamefont {E.}~\bibnamefont
  {Bennett}}, \bibinfo {author} {\bibfnamefont {G.}~\bibnamefont {Bergner}},
  \bibinfo {author} {\bibfnamefont {D.}~\bibnamefont {Elander}}, \bibinfo
  {author} {\bibfnamefont {C.~J.~D.}\ \bibnamefont {Lin}}, \bibinfo {author}
  {\bibfnamefont {B.}~\bibnamefont {Lucini}},\ and\ \bibinfo {author}
  {\bibfnamefont {M.}~\bibnamefont {Piai}},\ }\href
  {https://doi.org/10.1007/JHEP06(2016)114} {\bibfield  {journal} {\bibinfo
  {journal} {JHEP}\ }\textbf {\bibinfo {volume} {06}},\ \bibinfo {pages}
  {114}},\ \Eprint {https://arxiv.org/abs/1605.04258} {arXiv:1605.04258
  [hep-th]} \BibitemShut {NoStop}%
\end{thebibliography}%

\end{document}